%% file: stability_arxiv.tex
\documentclass[usenatbib]{mn2e}
\usepackage{graphicx}
\usepackage{amsmath}
\usepackage{amssymb}
\usepackage{abbv}
\usepackage{listings}
\usepackage{longtable, lscape} 
\newcommand{\bbcrit}{$\beta / \beta_{\rm crit}$}

\title[Super Earths and Dynamical Stability of Planetary Systems]{Super Earths and Dynamical Stability of Planetary Systems: First Parallel GPU Simulations Using GENGA}

\author[S. Elser, S.L.Grimm and J.G.Stadel]{S.~Elser,$^1$ S.~L.~Grimm,$^1$ J.~G.~Stadel$^1$ \\
$^1$Universit\"at Z\"urich, Winterthurerstrasse 190, CH-8057 Z\"urich, Switzerland}
\begin{document}

\maketitle

\begin{abstract}
We report on the stability of hypothetical Super-Earths in the habitable zone of known multi-planetary systems. Most of them have not yet been studied in detail concerning the existence of additional low-mass planets. The new N-body code GENGA developed at the UZH allows us to perform numerous N-body simulations in parallel on GPUs. With this numerical tool, we can study the stability of orbits of hypothetical planets in the semi-major axis and eccentricity parameter space in high resolution. Massless test particle simulations give good predictions on the extension of the stable region and show that HIP 14180 and HD 37124 do not provide stable orbits in the habitable zone. Based on these simulations, we carry out simulations of $10 M_{\oplus}$ planets in several systems (HD 11964, HD 47186, HD 147018, HD 163607, HD 168443, HD 187123, HD 190360, HD 217107 and HIP 57274). They provide more exact information about orbits at the location of mean motion resonances and at the edges of the stability zones. Beside the stability of orbits, we study the secular evolution of the planets to constrain probable locations of hypothetical planets. Assuming that planetary systems are in general closely packed, we find that apart from HD 168443, all of the systems can harbor $10\,M_{\oplus}$ planets in the habitable zone.   
\end{abstract}


\section{Introduction}

In the past two decades, numerous planetary system have been discovered \citep{Schneider11}. Most of those systems contain only a single known planet. Since \cite{Butler99} announced the discovery of the first multiple planet system around a normal star, many multiple planetary systems were discovered and confirmed \citep{Wright10}. Many more will follow in the next few years when a high percentage of the present Kepler candidate planets are going to be confirmed \citep{Borucki11aa}. There are planetary systems with up to 6 planet candidates \citep{Lissauer11, Tuomi13}. Both the Doppler spectroscopy and the detection via transit observations prefer massive, respectively, large planets close to the host star. The discovery of Earth-like planet candidates with respect to mass and size has just started thanks to the high precision spectrograph HARPS \citep{Pepe11,Dumusque12} or space missions like Kepler \citep{Borucki12,Fressin12}, whereas planets of several Earth-masses, so called Super-Earth, were discovered in the habitable zone of stars \citep*{Vogt12, LoCurto13}. Nevertheless, the detection of a Earth-like planets in the habitable zone around a Sun-like star is extremely difficult and was not yet successful.   

To guide the search for additional planets in known planetary systems, numerical stability studies are a powerful tool. In the recent years, numerical investigations estimated stability zones in known systems which might harbor unknown planets \citep{Menou03,Asghari04,Barnes04, Raymond05, Hinse08,Kopparapu08, Fang12}. \cite{Barnes04} and \cite{Raymond05} had shown the location of a stable zone in the 55 Cancri system before planet $f$ was discovered right at the inner edge of this zone \citep[e.g.][]{Fischer08}. They also predicted the existence of a Saturn-mass planet in HD 74156, which was later discovered by \cite{Bean08}. However, this prediction of the orbit and mass of an extra planet is not yet confirmed and under debate \citep{Baluev09,Wittenmyer09}.

Many multiple planetary systems tend to be near the edge of stability and small perturbations would destabilize the system \cite[e.g.][]{BarnesQuinn04}. The ``Packed Planetary Systems'' (PPS) hypothesis \citep{Barnes04} claims that every stable region between two neighboring (known) planets is occupied by an additional (unknown) planet. Hence, all planetary systems tend to form ``dynamically full'' and have no large gaps between the planets. Based on this hypothesis, stability regions that are identified in between known planets should potentially host additional planets. Most likely, those planets are not very massive and the impact of their perturbation on the known planet orbits might be smaller than the observational limit. Hence, they can not be deduced from residuals in current (Doppler spectroscopy) data. 

There is a major drawback when studying the stability regions in present day planetary system configurations, because we do not take into account the effects of potential early evolution of the known planets on the formation and evolution of the hypothetical planets. Despite this, early migration of giant planets through the initial planetesimal belt need not inhibit the formation of terrestrial planets, as long as the migration time scale is small \citep*[e.g.][]{Mandell03,Raymond06}. Dynamical instability of the initial giant planet configuration may result in ejection of one of the giants or in a merger with the central star. Such events might strongly affect the stability of a hypothetical Super-Earth sized planet located in the stability region of the final giant planet configuration and would also explain the high eccentricities of many of the observed planets. Hence, the width of stable regions in the parameter space are overestimated when dynamical instability played a significant part in forming the final giant planet configuration \citep*{Matsumura12}. However, if we assume that some hypothetical planets might form or survive despite of the early evolution of the known planets in a system, they would be found in the stability zones studied in this work.

The goal of this study is the prediction of stable orbits in the habitable zone of various extra solar multiple planetary systems, most of which have not yet been studied in much detail concerning stability of hypothetical planets. As a  major selection criterion, we chose systems whose inner- and outermost observed planets (partially) enclose the habitable zone of the system. To calculate the orbital movement of the planets, we use a new code developed at the UZH called GENGA (Grimm \& Stadel 2013, in preparation), which runs completely on a graphics processing unit (GPU). This simulation code allows either a single integration with many bodies (up to ten thousand massive bodies and hundreds of thousands of massless test particles), or many parallel integrations of systems with fewer bodies to be performed on a GPU. We start to constrain stable regions in the parameter space of semi-major axis and eccentricity of a hypothetical planet analytically based on the present planets orbits. This is the first indicator on the presence of a stable zone in the initial parameter space. Then, we integrate the orbits of massless test particles in the habitable zone of the planetary systems.  Finally, we focus on the identified stability regions and perform a large number of simulations to explore the parameter space in more detail. In this case, each simulation contains the known planets plus a massive hypothetical test planet. The stability of the test planet and its perturbations on the known planets indicate if a massive planet can be present in the habitable zone. All told, these simulations required around 2500 GPU-days or 2 months of wallclock time on our CPU-cluster.

This work is structured as follows: in section \ref{sec:datamethods}, we present the systems that we take into account. Moreover, analytic approaches to estimate the stability of a planetary system are briefly presented. Then, in section \ref{sec:simulations}, we introduce GENGA and show some comparisons with similar codes to highlight the advantages of this powerful new tool. In addition, we present the set up for the simulations with massless particles and massive hypothetical Super-Earths. Section \ref{sec:results} shows the main results. Besides presenting the extent of the stability region in each system, we highlight the most important insights and constrain the most likely regions where hypothetical Super-Earths may still be found. Finally, we conclude this work in section \ref{sec:conclusion}.


\section{Data and Methods}
\label{sec:datamethods}

First, our data sample is described and we explain our motivation to choose this set of systems. Then, the packed-planetary-systems hypothesis is briefly described. Analytic methods to predict stable orbit locations in the semi-major axis and eccentricity parameter space are shown. 

\subsection{Data sample selection}

The search for habitable planets is one of the main goals of present day astronomy. A habitable planet is often described as a terrestrial planet of the order of the mass of the Earth up to the mass of a Super-Earth ($\approx 10\,M_\oplus$) located in the habitable zone of its host star. The habitable zone (HZ) of a star is given by an annulus in distance where water on the surface of a planet can sustain in liquid form. A more general concept that takes into account the average time of a planet spending in the HZ is the eccentric habitable zone (EHZ). The exact definitions that are used in this work are given in appendix A. 

We focus on systems in which the HZ is enclosed between the orbits of the inner- and outermost planets. If the HZ is enclosed only partially, the enclosed fraction should be significant, that means more than half of the HZ. Otherwise, most planets initially located inside the HZ will be perturbed or crash with the known planet. Focusing on such systems with (partially) enclosed HZ, the parameter space of interest is limited by the planets in the system and its HZ. If the PPS-hypothesis holds, every stable zone we find should potentially harbor (at least) an additional planet as a consequence of the systems formation process. 
The sample we use is shown in table \ref{tab:systems}. Our sample does not represent all known multi-planetary systems with a (partially) enclosed HZ. In order to produce new results and save computational resources, we focus on systems that have not yet been studied in detail concerning stable region in the HZ (beside HD 47186, which allows a direct comparison of our simulation method). Hence, we exclude systems like 55 Cancri or HD 74156, which would also correspond to our selection criterion. In addition, we do not take into account any Kepler candidate systems.

\begin{table*}
\begin{minipage}{168mm}
\centering
\caption{Planetary systems of this study. The stellar mass ($M_\star$), the stellar surface temperature	($T_\star$) and the stellar radius ($R_\star$) are shown. For each planet in the system, the minimum mass ($	m \sin i$), the semi-major axis ($a$) and the eccentricity ($e$) are listed. Data from exoplanets.org \citep{Wright11exoplanets.org} in 2012 September 18.} 
\begin{tabular}{l c c c c c c c }
\hline
\input{planetlist.txt}
\end{tabular}
\label{tab:systems}
\end{minipage}
\end{table*}

\begin{table}
\centering
\caption{Values of \bbcrit of planetary systems in this studies. Systems with more than 2 known planets are marked with (a). In this case, the planet pair enclosing the HZ is taken into account. } 
\begin{tabular}{l c r}
\hline
\input{bvalues.txt}
\end{tabular}
\label{tab:bvalues}
\end{table}

\subsection{Analytic predictions}
\label{sec:analytic}
Before studying the planetary systems with numerical methods, we present some analytic approaches with various levels of complexity to constrain and to quantify the stability in a system. Although none of them can predict details on the stability region in the (a,e)-plane, they are by far less time consuming and are the first step when studying a system.
 
In the case of two-planet systems, an analytic stability boundary \citep{Barnes06,Barnes07} can be calculated, which is based on fundamental quantities of the system. Following \cite{Marchal82} and \cite{Gladman93}, the system is called Hill-stable and the orbits of the planets will never cross, if the ratio $\beta/\beta_{\rm crit}$ is larger then unity.  $\beta$ is a quantity that depends on the energy and the total angular momentum of the system, $\beta_{\rm crit}$ depends only on the masses of the star and planets:
\begin{equation}
\beta = \frac{-2(M_{\star} + M_{\rm 1} +M_{\rm 2})}{G^2(M_{\rm 1}M_{\rm 2}+M_{\star}M_{\rm 1}+M_{\star}M_{\rm 2})^3}L^2E
\label{eq:bbcrit}
\end{equation}
\begin{equation}
\beta_{\rm crit}=1+\frac{3^{4/3}M_{\rm 1}M_{\rm 2}}{M_{\star}^{2/3}(M_{\rm 1}+M_{\rm 2})^{4/3}}-\frac{M_{\rm 1}M_{\rm 2}(11M_{\rm 1}+7M_{\rm 2})}{3M_{\star}(M_{\rm 1}+M_{\rm 2})^2}+ ...,
\end{equation}
where $M_{\star}$, $M_{\rm 1}$ and $M_{\rm 2}$ are the masses of the star and two planets, given that $M_{\rm 1}>M_{\rm 2}$. Here $G$ is the gravitational constant and $E$ and $L$ are the total energy and orbital angular momentum of the system. This ratio, shown in table \ref{tab:bvalues} for each system, can be used to predict the possible existence of additional planets.  
According to \cite{Barnes07}, numerical simulations have shown that \bbcrit$\lesssim 1.5$ indicates that the system tends to be fully packed, whereas a system with \bbcrit$ \gtrsim 2$ offers stable zones for additional unknown planets. For $1.5\lesssim$ \bbcrit$\lesssim 2.0$, it is not clear if the system is packed or not. The four systems that contain more than 2 known planets are also listed. It is not guaranteed that \bbcrit=1 means Hill stability of any individual pair. The above limits hold if the additional planets in the system are small (e.g. HIP 57274) or well separated compared to the pair that is taken into account for calculating \bbcrit. Based on the above argument, we expect stable orbits in all systems apart from HIP 14180 and HD 37124.

The main osculating elements of the known planets constrain the osculating elements of any hypothetical planet.
A test particle whose initial orbit crosses that of a planet is highly in danger of colliding or being scattered out of the system as a result of a close encounter. The location of crossing orbits are given by the point in the (a,e)-plane where pericenter (resp. apocenter) and apocenter  (resp. pericenter) of a particle and a planet coincide. These limits provide a very general constraint on the size and shape of the stability region in the (a,e)-plane. 

Capture in low order mean motion resonance (MMRs) can provide stability beyond the crossing orbit of the planets, \cite[e.g.][]{Kopparapu08,Raymond08}, or destabilize planets in the stability region. More important, the zone of the dynamical influence of a planet is larger than its physical cross section. This gravitational zone of influence of a planet $i$ is often expressed as some factor $c_i$ times the Hill radius \citep{Hamilton92},
\begin{equation}
R_{\rm Hill,i} \equiv \left(\frac{M_{p,i}}{3M_\star}\right)^{1/3} a_i,
\end{equation}
where $i=1,2$ refers to the enclosing planets. 
Without loss of generality, we take into account a two planet system where $(a_1,e_1)$ are the osculating elements of the inner planet and $(a_2,e_2)$ the corresponding elements of the outer planet. The lines of crossing orbits in (a,e)-space are given by
\begin{equation}
a_1(1+e_1)+c_1R_{\rm Hill,1}=a(1-e),
\label{eq:limits1}
\end{equation}
\begin{equation}
a(1+e)=a_2(1-e_2)-c_2R_{\rm Hill,2}.
\label{eq:limits2}
\end{equation}
In general, the factors $c_i$ are unknown and $c_1= c_2=0$ provides a first insight. To account for the dynamical influence, a common choice is $c_1= c_2=3$ \citep{Menou03} or higher. Studying Kepler systems with two known planets, \cite{Fang12} obtained $c_1=19.4$ for accounting the influence of the inner planet outwards and $c_2=4.2$ for accounting the influence of the outer planet inwards. \cite*{Jones06} used cubic fits on $c_1$ and $c_2$ obtained from simulations to get the factors for any planetary system by interpolation. They found $1\lesssim  c_2\lesssim  3$, decreasing with planet eccentricity and $3\lesssim  c_1\lesssim 13$, increasing with planet eccentricity. For our purpose, $c_1$ and $c_2$ are estimated by solving the system of equations (\ref{eq:limits1}) and (\ref{eq:limits2}) for $e$ to get a piecewise function $e=e(a,c_1,c_2)$. Then, this function is fitted to the edge of the stable regions in the (a,e)-plane, which gives $c_1$ and $c_2$. The maximum eccentricity of all stable particles, $e_{\rm top}$, is then given by the maximum of the function $e=e(a,c_1,c_2)$ and can be interpreted as a measure of the stable zone. We use this to estimate $e_{\rm top}$ and to check how well the edges of the stability region can be expressed by equations (\ref{eq:limits1}) and (\ref{eq:limits2}). 

Since analytic estimates are limited and e.g. the estimation of the correct $c_i$ depends on numerical studies, we directly focus on N-body simulations to find stable orbits.


\section{Simulations}
\label{sec:simulations}

Similar to \cite{Raymond08} we use the term ``test planets'' for massive bodies which fully interact with the planets in contrast to the massless ``test particles'' which trace only the gravitational potential of the planets. To test the stability of planets, in a first attempt we used massless test particles which are computationally less expensive than simulations with massive test planets. 

The main orbital elements of the current best-fit orbits of the known planets are given in table \ref{tab:systems}. The minimum mass, semi-major axis $a$ and eccentricity $e$ are shown with their observational uncertainties, which we use in the further study of the results. We randomly choose a mutual inclination of $i<1^{\circ}$ and a longitude of the ascending node randomly distributed from $0^{\circ}$ to $360^{\circ}$. The argument of periastron and the time of periastron passage are given by the references. If the actual inclination of the system were larger than a few degrees, the planet masses would be much larger and would change the dynamics of the system significantly.

In every planetary system, the two planets enclosing the HZ are named as follows: the planet whose semi-major axis is smaller than the center of the HZ (\ref{eq:HZparas}) is called the ``inner'' planet, the planet whose semi-major axis is larger than the center of the HZ is called ``outer'' planet.  

In each simulation, the goal is to conserve energy up to 1 part in $10^5$. This is achieved by choosing a suitable combination of time step and order of the integrator for each system, table \ref{tab:percentages}. The maximum time step is preset to 2 day.   

\subsection{The GPU Code GENGA}

Modern graphics cards and the specialized variants for pure computing found in supercomputers such as the CRAY-XK series can perform a large number of operations in parallel by launching a large number of execution threads. 
The limitation is that these threads are not independent and should perform, as much as possible, the same instruction on different data (SIMD). This type of high performance computing based on graphics processing units (GPU computing) can speed up many numerical tasks by a large factor over what is possible on a CPU as long as enough parallel work is available.
The simulation of planetary systems would seem to provide enough parallelism as long as enough bodies are involved in the simulation ($\gtrapprox$ 100) or enough independent systems are evolved simultaneously. Since the memory transfer between the CPU and GPU is currently still a bottleneck, GENGA runs completely on the GPU where it can take advantage of the very fast, but limited, memory that exists there. Only the outputs are transferred back to the CPU. GENGA is implemented in Cuda C by Grimm and Stadel and runs on NVidia GPUs with compute capability 2.0 or higher. A detailed paper is in preparation 
, but we will briefly present a few aspects of this new code in the following.

The GENGA Code is a hybrid symplectic integrator, based on the Mercury code \citep{Chambers98}. The symplectic integrator is a mixed variable integrator as described by \cite{Wisdom91,SahaTremaine92}. It integrates the planetary orbits for a large time scale with a very good energy conservation. Gravitational interactions between planets are computed as perturbations of the Keplerian orbits. If two planets are in a close encounter, the perturbation potential becomes dominant and the integrator breaks down. The hybrid symplectic integrator switches in these cases smoothly to a direct N-body Bulirsch-Stoer integrator which integrates the close encounter phase up to machine precision. Two planets are in a close encounter when their separation $r_{ij}$ is less than a critical radius, defined as 
\begin{equation}
r_{\rm crit} = \max(r_{\textrm{crit},i}, r_{\textrm{crit},j}),
\end{equation}
with
\begin{equation}
r_{\textrm{crit},i} = \max(3 \cdot R_{\textrm{Hill},i}, 0.4 \cdot \tau v_{\rm max}),
\end{equation}
where $\tau$ is the time step. In the GENGA code we generalized the second order symplectic integrator to fourth and sixth order, as described by \cite*{Yoshida90}. The higher orders are especially a good choice if the innermost planet has a very small semi-major axis and a high eccentricity. 

\begin{figure*} 
\centering
\includegraphics[width=0.7\textwidth]{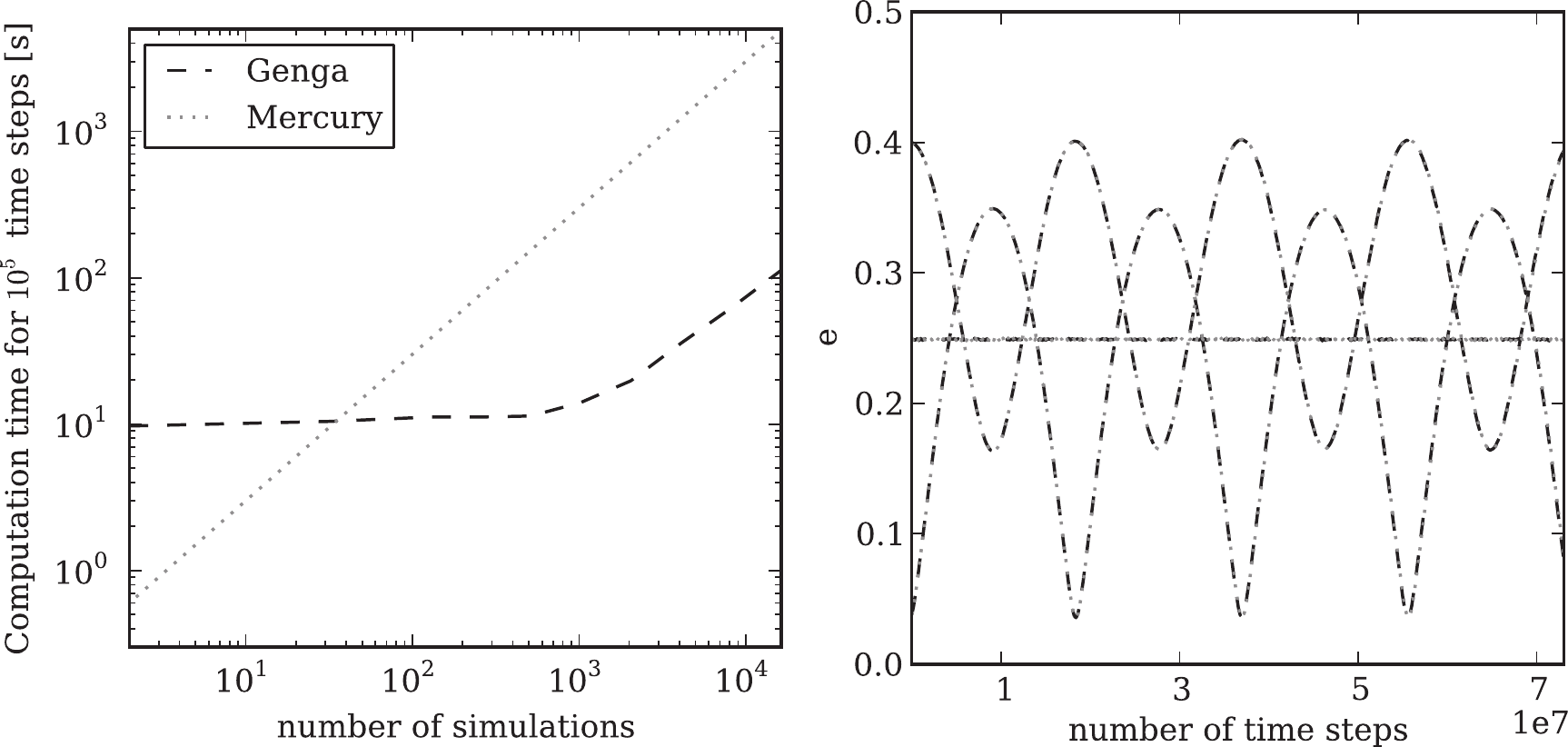} 
\caption{Performance of the GENGA Code. Left panel: Comparison of the performance between GENGA on one GPU (dashed line) and Mercury on one CPU (dotted) line. Right panel: Comparison of a simulation output. The secular evolution of the eccentricities of a three planet system is shown. It is HD 47186 with a $10\,m_\oplus$ test planet initially located at $(a,e)=(0.2\,\rm AU,0.4)$. Mercury and GENGA are in nearly perfect agreement here.}
\label{fig:M_Timing2} 
\end{figure*}

We use the GENGA Code in two different modes: First, to simulate the planetary systems with a large number of test particles, and second,
to simulate a large number of small, independent, planetary systems with different configurations. In the test particle mode we use one Cuda thread per test particle, in the multi simulation mode we use one Cuda thread per body. Figure\,\ref{fig:M_Timing2} shows the computation time for GENGA (on a NVidia Geforce GTX 590 graphic card) and Mercury (on an Intel Xeon 2.8\,GHz CPU) to simulate a set of three Body simulations. At a low number of simulations, the GPU overhead dominates and Mercury is faster. At a high number of simulations, GENGA benefits from the large number of GPU cores. At around 1000 simulations, the GPU is fully occupied, and the computation time begins to increase. At 16000 simulations the GPU is about 40 times faster than one CPU. 

The massive test planet simulations of each system are split onto 4 GPUs in most cases. This results in 1250 simulations per GPU, which allows the maximum efficiency of the code. The computation time depends on different factors: integration time step and order of the symplectic integrator, mainly controlled by the innermost planet, the survival rate of test planets and the number of close encounters. The minimum wallclock time is around 120 days for HD 147018 and the maximum wallclock time is around 800 days for HD 47186. 

\subsection{Massless test particle simulations}

In the test particle simulation, we placed 20'000 test particles equally spaced in 500 steps between the semi-major axis of the inner planet $a_{\rm inner}$ and the semi-major axis of the outer planet $a_{\rm outer}$ and equally spaced in 40 steps in $0.0\le e\le0.8$. The inclinations are assigned randomly under the condition $i<1^{\circ}$. The argument of periastron, the longitude of the ascending node and the mean anomaly are drawn randomly between $0^{\circ}$ and $360^{\circ}$.

A test particle is representing an unstable orbit if it collides with one of the known planets or if its distance to the star exceeds 20 AU. We stop the simulations when the overall shape of the orbital zone is visible, this means when the rate of orbits becoming unstable decreases significantly. This takes place after a few Myr and gives a rough idea of the stable zone and its features. Hence, we define stability by the survival of a planet. 

\subsection{Massive planets}

The initial sampling of the (a,e)-space in the case of massive test planets is guided by the results in the massless test particle simulations. The minimum and maximum $a$ and the maximum $e$ of the surviving test particles are approximately taken as limits. The extent of the sampled regions and further simulation details are given in table \ref{tab:percentages}. We run each simulation for $10\,$Myr. Most of the unstable orbits will be identified in $10^6$ orbits \citep{Barnes04}. The conditions for an orbit to be identified as unstable are the same as in the test particle simulations discussed previously.

Simulations with massive test planets provide additional information about the stability of planets in the system. Depending on the mass of the test planet, the orbital parameters of the other planets (and the star) might change due to secular interactions or close encounters. This can be used to narrow down possible orbits of the test planet \citep{Raymond08, Kopparapu08}. The ``fraction of time on detected orbits'' (FTD) quantifies the probability that the inner and outer planet are located at their observed best-fitted orbits, inside of the observational error bars. The back-reaction of the detected planets might be strong enough so that they spend a significant time outside the ($a,e$)-region they are observed in. Hence, the smaller the FTD the more unlikely the presence of a hypothetical planet on the corresponding initial orbit. This method is only applicable to systems were the secular interactions between the detected planets are small. Otherwise, the osculating elements $a$ and $e$ of the detected planet oscillate beyond their accredited orbits periodically without influence of a hypothetical planet \citep{Veras09}. Hence, we do not apply the FTD as an absolute constrain and only use it for planets which do not leave their observed ($a,e$)-region despite secular interaction with other observed planets in the system. 

Secular interaction among planets is a well studied field. The Lagrange-Laplace secular evolution theory, well described in \cite{Murray99}, allows to predict the long term evolution of eccentricity and inclination in multi-planet systems. The secular perturbation of the orbital elements are than given by the disturbing function expanded to second order in eccentricity and inclination. Thus, this classical theory demands that eccentricities and inclinations are small enough to guarantee that such an expansion is adequate. While all our simulations start with a small inclination, the eccentricities are sometimes rather large. However, since we use secular theory only as qualitative guideline to check the simulation results, it is not necessary to use higher order secular solutions (e.g \cite{Veras07}). 

Here, we apply the secular theory to calculate the effect of a known two-planet system on the hypothetical (massless) Super-Earth, following \cite{Adams06}. This holds for a massless particle, but it might hold also for Super-Earths, since the known planets in the systems are often much larger. With secular theory, the forced eccentricity component of a test particle can be calculated as a function of semi-major axis and time. Secular theory shows that the osculating eccentricity $e$ of a particle is composed of the time-dependent forced eccentricity $e_{\rm forced}$ and the free eccentricity $e_{\rm free}$ \citep{Murray99}. While the forced eccentricity is caused by the secular interactions with the known planets, the free eccentricity is basically given by the boundary conditions. The maximum value of $e$ is given by $ e_{\rm forced}+ e_{\rm free}$, the minimum is given by $\left| e_{\rm forced}- e_{\rm free}\right|$. If $e_{\rm forced}> e_{\rm free}$, particle oscillates around $e_{\rm forced}$ with amplitude $e_{\rm free}$. Otherwise, it oscillates around $e_{\rm free}$ with amplitude $e_{\rm forced}$. 

Most of the systems we study harbor planets on non-circular orbits. As mentioned above, secular interactions will force the orbits of neighboring test planets to change in eccentricity. On the other hand, MMRs or close encounters can cause a change in semi-major axis. To record the actual location of the test planets during the simulations, the ($a,e$)-plane is divided in multiple bins. The number of massive planets located in this bin in all simulations is summed over all time steps. Binning the presence of a stable particle in the ($a,e$)-plane results in the time-averaged location of all particles. It reveals the most likely eccentricity of a hypothetical planet for a given semi-major axis when it will be observed. 


\section{Results}
\label{sec:results}

\begin{table*}
\begin{minipage}{171mm}
\centering
\caption{The massive testplanet simulations in detail. The time step $\Delta t$ and the order of the integrator $\mathcal O$ are two parameters that control the accuracy of the simulation. The number of $N_{\rm init}$ simulations are sampled equally spaced in the (a,e)-plane in $a_{\rm min}\le a \le a_{\rm max}$ and $0\le e\le e_{\rm max}$. $N_{\rm stab}$ is the number of test planets that are on a stable orbit. $N_{\rm Hill}$ is the number of planets on a stable orbit that experience a close encounter. $f_{\rm stab}$ is the percentage of stable simulations in the massive test planet simulations. $F_{\rm stab}$ normalizes $f_{\rm stab}$ to the area between the planets and $0<e<1$. $F_{\rm stab,m=0}$ is the normalized percentage in the massless test particle simulations.}  
\begin{tabular}{l l l l l l l r r r r r }
\hline
\input{percentages.txt}
\end{tabular}
\label{tab:percentages}
\end{minipage}
\end{table*}

\begin{figure*}
\centering
\begin{minipage}{171mm}
\includegraphics[width=\textwidth]{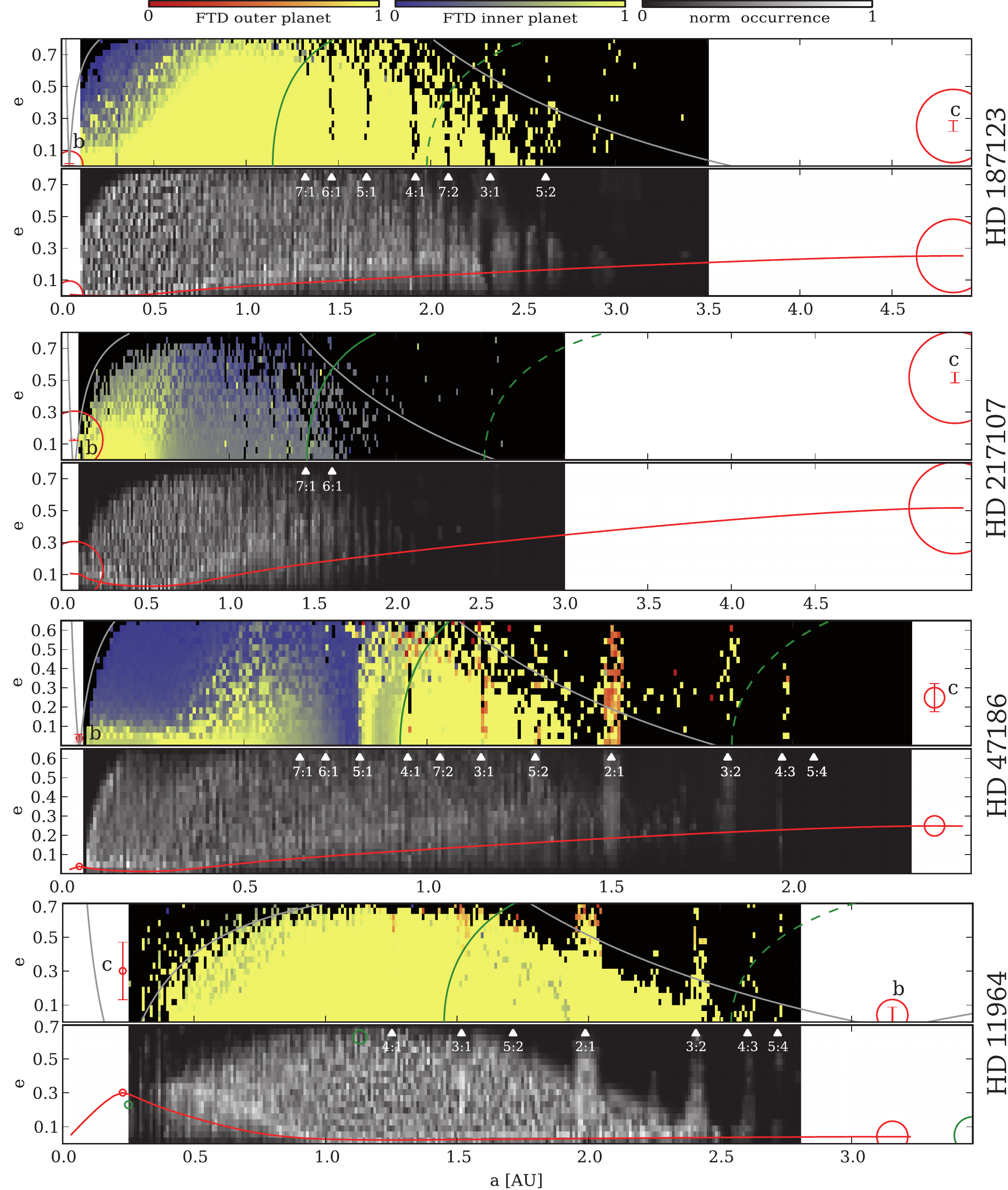}
\caption{Results of the massive test planet simulations in the systems HD 187123,  HD 217107, HD 47186 and HD 11964 with deceasing \bbcrit=(15.09,8.94,6.13,2.04). For each system, the results are presented in two panels. Top panel: The yellow region represents the orbital elements of massive test planets which were stable for 10 Myr. The black regions represents unstable regions. The color gradient from yellow to red represents orbits with a strong interaction with the inner planet, this means the fraction of time on detected orbit (FTD) decreases. The gradient from yellow to blue represents orbits with a strong interaction with the outer planet (here planet $d$). The gray lines show the location of the crossing orbit of the planets. The full green lines gives the inner edge of the EHZ, the dashed green line gives the outer edge. Bottom panel: The occurrence of a test planet in a given parameter space bin during the whole simulation normalized to 1. The brighter the color the more likely is it to observe a planet with orbital elements according to this bin. The red line gives the value of the forced eccentricity due to secular perturbation. The location of the most important MMRs is also shown. The green circles in the lower panel of HD 11964 shows the three planet solution by \protect \cite{Gregory07}.}
\label{fig:panel1}%
\end{minipage}
\end{figure*}

\begin{figure*}
\centering
\begin{minipage}{171mm}
\includegraphics[width=\textwidth]{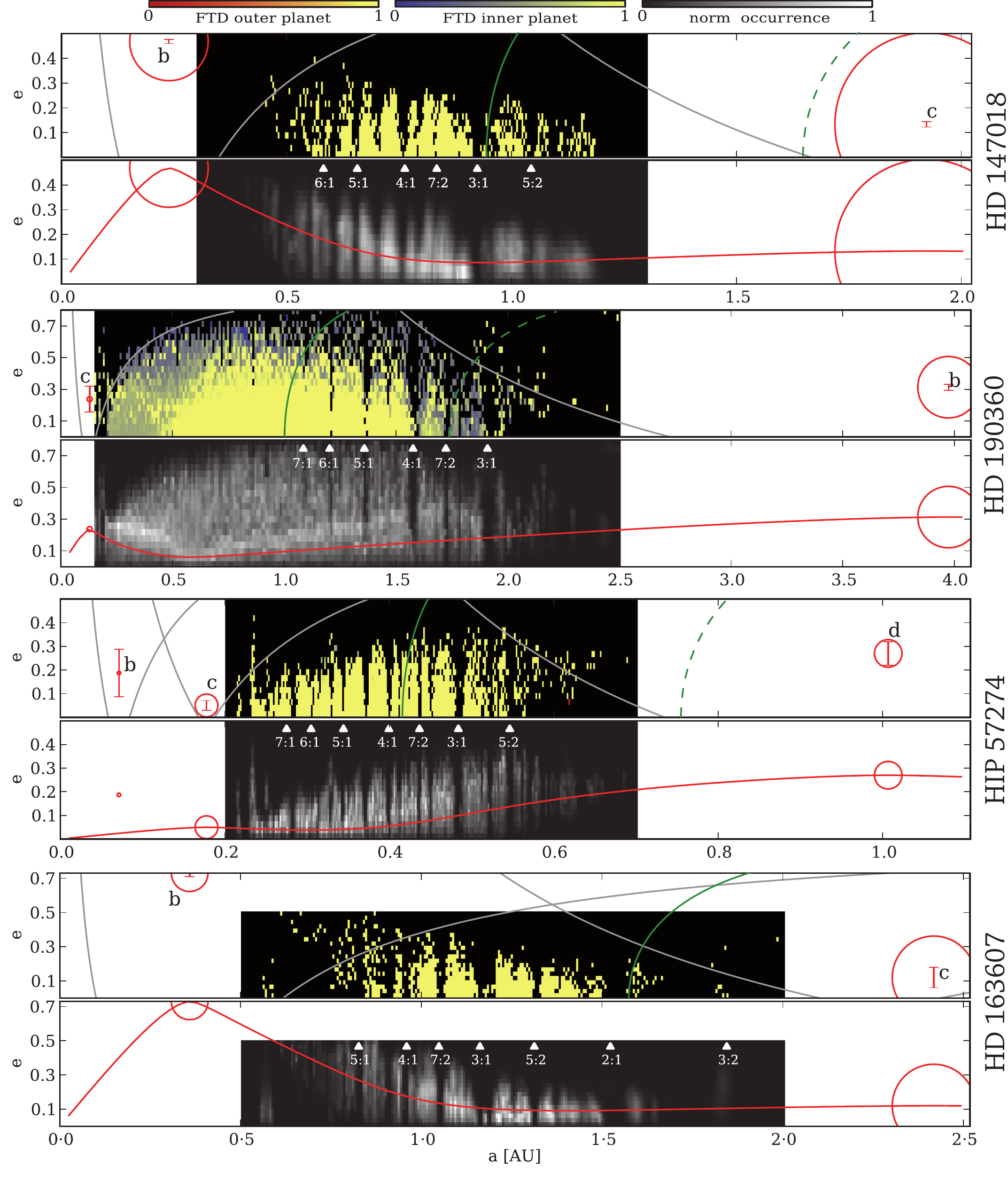}
\caption{Results of the massive test planet simulations in the systems HD 147108, HD 190360, HIP 57274 and HD 163607 with deceasing \bbcrit=(2.01,1.80,1.78,1.58). For each system, the results are presented in two panels. Top panel: The yellow region represents the orbital elements of massive test planets which were stable for 10 Myr. The black regions represents unstable regions. The color gradient from yellow to red represents orbits with a strong interaction with the inner planet, this means the fraction of time on detected orbit (FTD) decreases. The gradient from yellow to blue represents orbits with a strong interaction with the outer planet (here planet $d$). The gray lines show the location of the crossing orbit of the planets. The full green lines gives the inner edge of the EHZ, the dashed green line gives the outer edge. Bottom panel: The occurrence of a test planet in a given parameter space bin during the whole simulation normalized to 1. The brighter the color the more likely is it to observe a planet with orbital elements according to this bin. The red line gives the value of the forced eccentricity due to secular perturbation. The location of the most important MMRs is also shown.}
\label{fig:panel2}%
\end{minipage}
\end{figure*}

\begin{figure*} 
\centering
\begin{minipage}{171mm}
\includegraphics[width=\textwidth]{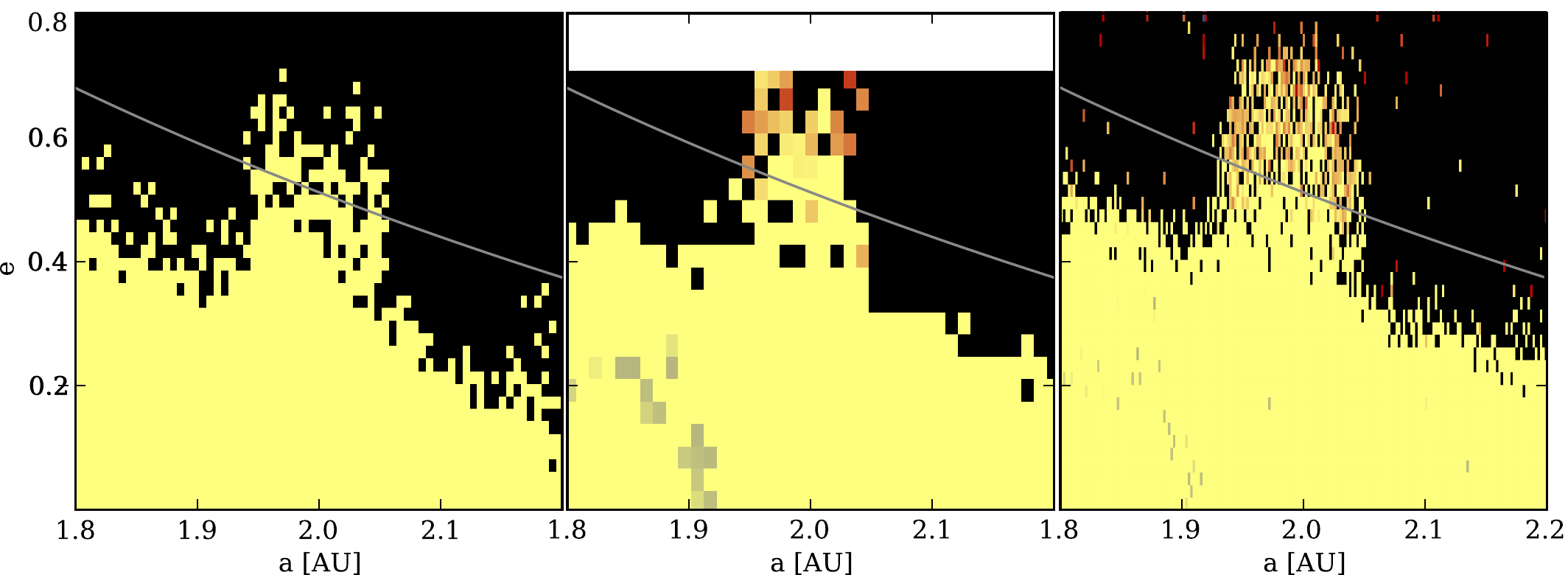}
\caption{$2d:1c$ MMR of the HD 11964 system. The left panel shows a detail of the test planet simulation. The central panel shows a detail of the massive test planet simulations presented in figure \ref{fig:panel1} ($e\leq0.7$). The right panel shows massive test planets simulations carried out in higher resolution (200$\times$40 simulations).  The color gradient is given in figure \ref{fig:panel1}. Particles and test planets initially located close to the resonance ($\pm 0.05\,$AU) become stable. If they are located above the line of crossing orbits, they significantly diminish the FTD of planet $b$. }
\label{fig:zoom}%
\end{minipage}
\end{figure*}

\begin{figure*} 
\centering
\begin{minipage}{171mm}
\includegraphics[width=\textwidth]{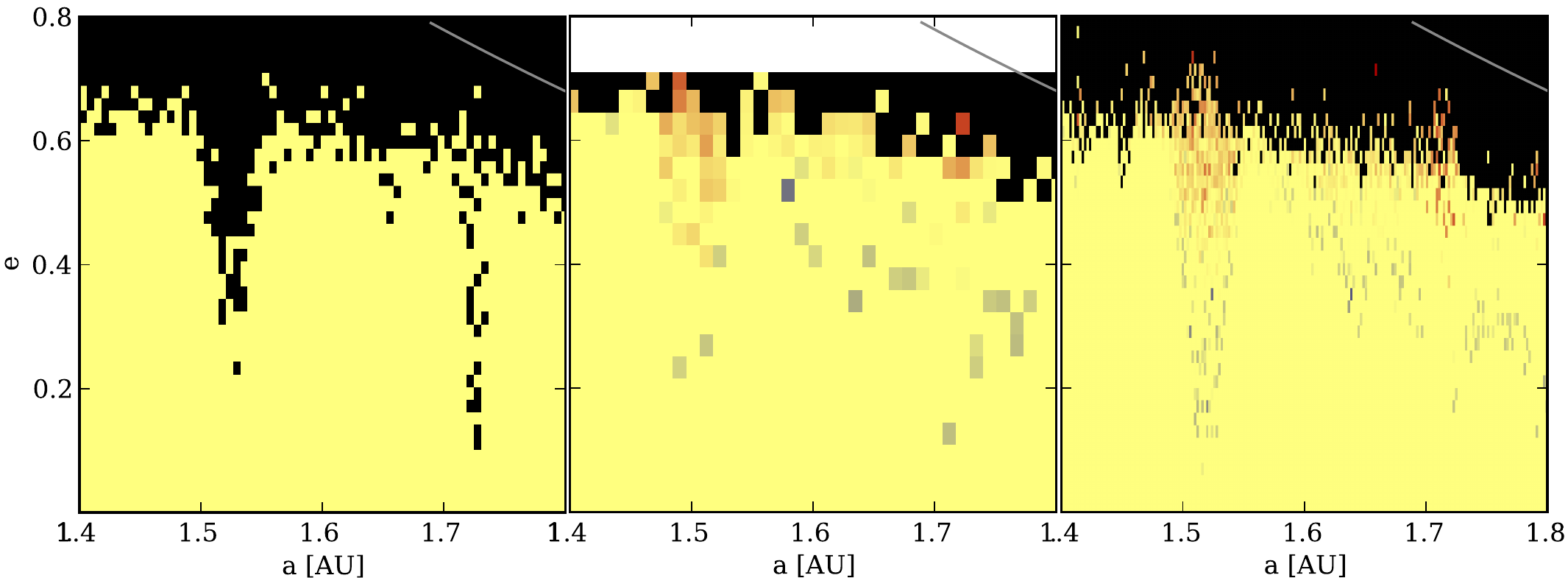}
\caption{$3d:1c$ and $5d:2c$ MMRs of the HD 11964. The left panel shows a detail of the test planet simulation. The central panel shows a detail of the massive test planet simulations presented in figure \ref{fig:panel1} ($e\leq0.7$). The right panel shows massive test planets simulations carried out in higher resolution (200$\times$40 simulations). While massless particle in MMR with the outer planet become unstable, the massive test planets are stable. They diminish mostly the FTD of the outer planet.}
\label{fig:zoom2}%
\end{minipage}
\end{figure*}

The massless test particle simulations reveal that not all systems are worth further detailed study. They show that the HIP 14180 triple giant plant system harbors test particles in a well defined region in between the two inner planets. Between the two outer planets, where the HZ is located, only very few orbits are stable. Hence, we do not perform additional simulations with massive test planets. HD 37124 hosts three giant planets of almost equal mass. The inner edge of the HZ coincides with the apocenter of the inner most planet. As a result of the relatively high masses of the planets and their non-zero eccentricity, all test particles are lost in a few 100'000 years and we do not carry out the simulations with massive test planets.

Finally, we focus on the eight systems that are most likely to provide stable orbits in the EHZ. Hence, we carry out massive test planet simulations for the systems HD 11964, HD 47186, HD 147018, HD 163607, HD 187123, HD 190360, HD 217107 and HIP57274. The main results are given in table \ref{tab:percentages} and figures \ref{fig:panel1} and \ref{fig:panel2}. They show the location of the stable orbits of 10 $M_\oplus$ mass planets in the systems, given that the orbital solution for the known planets is correct. HD 168443 hosts two known companions: the inner one is a very massive giant planet, the outer a brown dwarf.
Test particle simulations reveal that some stable orbit exist around 1 AU at low eccentricities. Although this is not part of the EHZ, we carry out the massive test planet simulations to check if massive Super-Earths may survive. Only very few planets are stable. Hence, this systems is not shown as a figure.

In table \ref{tab:percentages} the fractions of orbits that are stable ($f_{\rm stab}$) are listed. The normalized fractions $F_{\rm stab}$ are given by the percentage of the area in $a_{\rm inner}<a<a_{\rm outer}$ and $0.0<e<1.0$ that is covered by the stable orbits. 

Based on the numerous massive test planet simulations that are carried out, we present the major insights in the following subsection.

\subsection{Zones of dynamical influence} 
The lines of crossing orbits give a fundamental constraint on the stability regions. In addition to the physical cross section, dynamical interaction plays a major role. In HD 11964, HD 47186, HD 187123, HD 190360 and HD 217107 the shape of the stability region is clearly following the lines of crossing orbits with a partially significant offset. Towards the inner planet, the line traces the outer edge very well, apart from high $e$. The outer edge of the stability zone is shifted inwards due to the dynamical influence of the outer planet. The relatively large semi-major axis of this planet results in a large Hill-radius and dynamical influence. In addition, higher eccentricity of the outer planet leads to a more diffuse transit from stability to instability, in our examples often in combination with MMRs. In contrast, low eccentricity results in a sharp edge (HD 11964). 

In the case of HIP57274, HD 163607 and HD 147018, which are systems with strong interaction among the planets, the stability regions are significantly truncated compared to the line of crossing orbits. Beside the large masses of the inner planets, their relatively large semi-major axes enhance their dynamical influence. In addition, the dominant MMRs of the outer planets amplify this effect. Secular resonances result in oscillation of the eccentricity of the known planets. Therefore, the lines of crossing orbits change on a secular time scale. Nevertheless, the shift of the lines of crossing orbits due to the oscillations is too small to truncate the stable region additionally over time.

\subsection{Significant MMRs}
The MMRs play a major role in shaping the stable regions of many systems. In most of the systems the outer planet has a relatively high eccentricity ($e>0.1$) and mass ($m>1.5\,M_{\rm Jupiter}$). The MMRs of this planet with the test planet tend to destabilise the later. This cuts narrow wedges into the outer part of stability zone (e.g. HD 187123, HD 190360 or HD 217107). Since they can be very narrow, their visibility is sometimes limited by the finite resolution of our sampling. In combination with a highly eccentric ($e>0.4$) and massive inner planet, the stability region tends to be completely divided by MMRs, because the orbit of the inner planet truncates the stable zone significantly. 

In contrast, small planets ($m<1.0\,M_{\rm Jupiter}$) with low eccentricity ($e<0.05$) can provide additional stable zones due to MMRs, because their dynamical influence is not that strong. In the region beyond the limits of crossing orbits, test planets can be captured by the strong MMRs (HD 11964: $2d:1c$ and $3d:2c$, HD 47186: $2d:1c$). These MMRs tend to catch particles which would be potentially unstable. In HD 11964 the high order MMRs ($3d:1c$) are not strong enough to cut into the stable zone. To exclude a resolution effect, a high resolution zoom in the parameter space around the $2d:1c$ and $3d:1c$ MMRs with 8000 simulations was calculated and is shown in figure \ref{fig:zoom}. The location of the $3d:1b$ and $5d:2b$ MMRs in the massless particle simulations are cleaned, while in the massive planet simulations, the orbits captured in the MMRs are stable. A decrease in the FTD of planet $b$ at the MMRs indicates a strong interaction among the planets in resonance.  

\subsection{Fraction of time on detected orbits}

The FTD is diminished in large regions of the stability zone in many systems. The inner planet is often perturbed significantly by the test planets. Typically, test planets close to the inner planet play a major role. The more the initial eccentricity coincides with the initial eccentricity of the inner planet, the smaller is their effect (e.g. HD 47186, HD 187123, HD 190360). If the eccentricities do not coincide, the eccentricity of the inner planet is forced to change. MMRs are the only occasions where the FTD of the outer planet is diminished (HD 11964, HD 47186). In this case, the resonance with the test planet is strong enough to perturb the outer planet with $m<1\,M_{\rm Jupiter}$ significantly. In the zoom simulation of two details of HD 11964 (figure \ref{fig:zoom} and \ref{fig:zoom2}), this effect is clearly visible. 

An interesting feature can be observed in some of the systems: The FTD of the inner planet has a minimum in parts of the stability region while between this minimum and the inner planet, at the same eccentricity, the FTD reaches 1. This is observed in HD 47186, HD 187123, HD 190360, HD 217107 and marginally in HD 11964. This effect is caused by secular resonances and depends very much on the architecture of the system and on the given error bars. An illustrative example is given by HD 47186. The simulations show that there is no continuous region of high FTD at $a<0.9\,$AU.  In fact, a minimum in the FTD around 0.7 AU of FTD$\approx0.3$ with respect to the inner planet is found for all eccentricities. Secular perturbations of the outer planet let the test planet oscillate according to the corresponding $e_{\rm free}$ and $e_{\rm forced}$. This results in the eccentricity oscillation of the inner planet which reacts significantly due to its relatively small mass of around 22 $M_\oplus$. One can say that the test planet acts to transfer a secular perturbation from the outer planet onto the inner one. If the test planet is located closer to the inner planet, $e_{\rm forced}$ is smaller. Therefore, its secular oscillation is too small to affect the inner planets FTD. If the test planet is further away from the inner planet, its forced oscillation can hardly be transferred to the inner planet. 

HD 47186 was already studied in detail with lower resolution by \cite{Kopparapu08}. Our stability region agrees with their results, but the FTD results differ. \cite{Kopparapu08} found a sharp border in the FTD at $a\approx0.25$ dividing a region of very low ($\approx0.2$) FTD and the broad region of FTD=1 between 0.3 and 1.3 AU. We found out that this disagreement with \cite{Kopparapu08} is caused by different time steps used in the integration. Secular oscillations of the planets' eccentricities are sometimes missed in \cite{Kopparapu08}(private communication).

Regarding the existence of possible orbits in the EHZ, the FTD provides significant constrains only in the case of HD 217107. This is a result of the average location of the HZ, whose distance to the inner planet is often large and resulting secular perturbations are small.

The FTDs of the known planets were not studied in the case of the planets in HD 163607 and HD 147186 and planet $c$ in HIP 57274. They were excluded because of strong secular perturbations among the known planets. 

\subsection{Massless test particles}
The massless test particle simulations provide very detailed pictures of the stability regions. Comparisons of the area of the stable zone found in the massless test particle simulations and results of the massive test planet simulations show that both are very similar. The normalized percentages of stable orbits are listed in table \ref{tab:percentages}. The most significant difference is prominently seen in HD 11964. In figure \ref{fig:zoom}, a detailed comparison with a the test particle simulation, the low resolution and the high resolution simulation set of massive test planets is shown. The location of the $4d:1b$ and $3d:1b$ MMRs in the massless particle simulations are cleaned, whereas in the massive planet simulations, the planets in the MMRS are stable.  Obviously, the mass of the test planet adds additional stability to the MMRs. Beside the MMRs, the low and high resolution simulations with massive test planets agree very well with the massless particle simulations. In some parts, it seems that the test particle simulation can not reproduce the complete area of stable orbits at the very edge of the stability region. Beside the effect of lower resolution, a possible explanation is that test planets involved in close encounters are not as much perturbed as massless particles. 

\subsection{Forced eccentricity}
The lower panel of each system's plot (Figures \ref{fig:panel1} \& \ref{fig:panel2}) shows the normalized occurrence rate. It gives the time-averaged location of all stable orbits and represents the likelihood that a hypothetical planet is found in a certain bin of the (a,e)-plane. Many systems show a prominent curve of maximum occurrence rate (e.g. HD 190360, HD 47186). The curves approach asymptotically the eccentricity of the inner and outer planets, often with a minimum in eccentricity. This shows that the test planets are forced to change their eccentricity.

When comparing the analytically estimated amplitude of $e_{\rm forced}$ given by secular theory and the most likely location of the test planets in the (a,e)-plane, the minimum of the predicted forced eccentricity clearly coincides with the minimum of the curve in the occurrence rate. The maximum of the occurrence rate along the eccentricity does not agree with $e_{\rm forced}$. Beside the limitations of the secular theory at high eccentricities, this is caused by the fact that the particle oscillates around $e_{\rm forced}$ only when $e_{\rm forced}>e_{\rm free}$. Hence, the planets with initially high eccentricity tend to spend most of their time at high $e$. Therefore, we have to point out that the measured occurrence rate depends on the expansion of the sampled region along the $e$-axis. 

Although the averaged flux that the planet receives is more important for habitability than the planet's eccentricity (appendix A), a small $e_{\rm forced}$ can be interpreted as a optimal location for a habitable planet, following \cite{Adams06}. When we assume that the particle is initially on a low eccentric orbit, the $e_{\rm forced}$ gives the more realistic eccentricity than the occurrence rate.

\subsection{Close encounters}  
The numbers of stable test planets that were part of a close encounter are given in table \ref{tab:percentages}. The fraction of such stable orbits is $\approx15$ per cent in HD 147018 or $\approx4$ per cent in HD 11964. (In HD 168443, almost all stable test planets had a close encounter but since only $\approx0.1$ per cent of all configurations are stable, this is not surprising.) Most close encounters take place at the outer edge of the stability region. This confirms our decision not to classify an orbit as unstable as soon as its planet has a close encounter. Thus, the criterion to identify unstable orbits should not be given by the occurrence of a close encounters.  Nevertheless, there are systems where no close encounters of the stable test planets take place.

\subsection{Analytic predictions}

\begin{figure}
\centering
\includegraphics[scale=0.8]{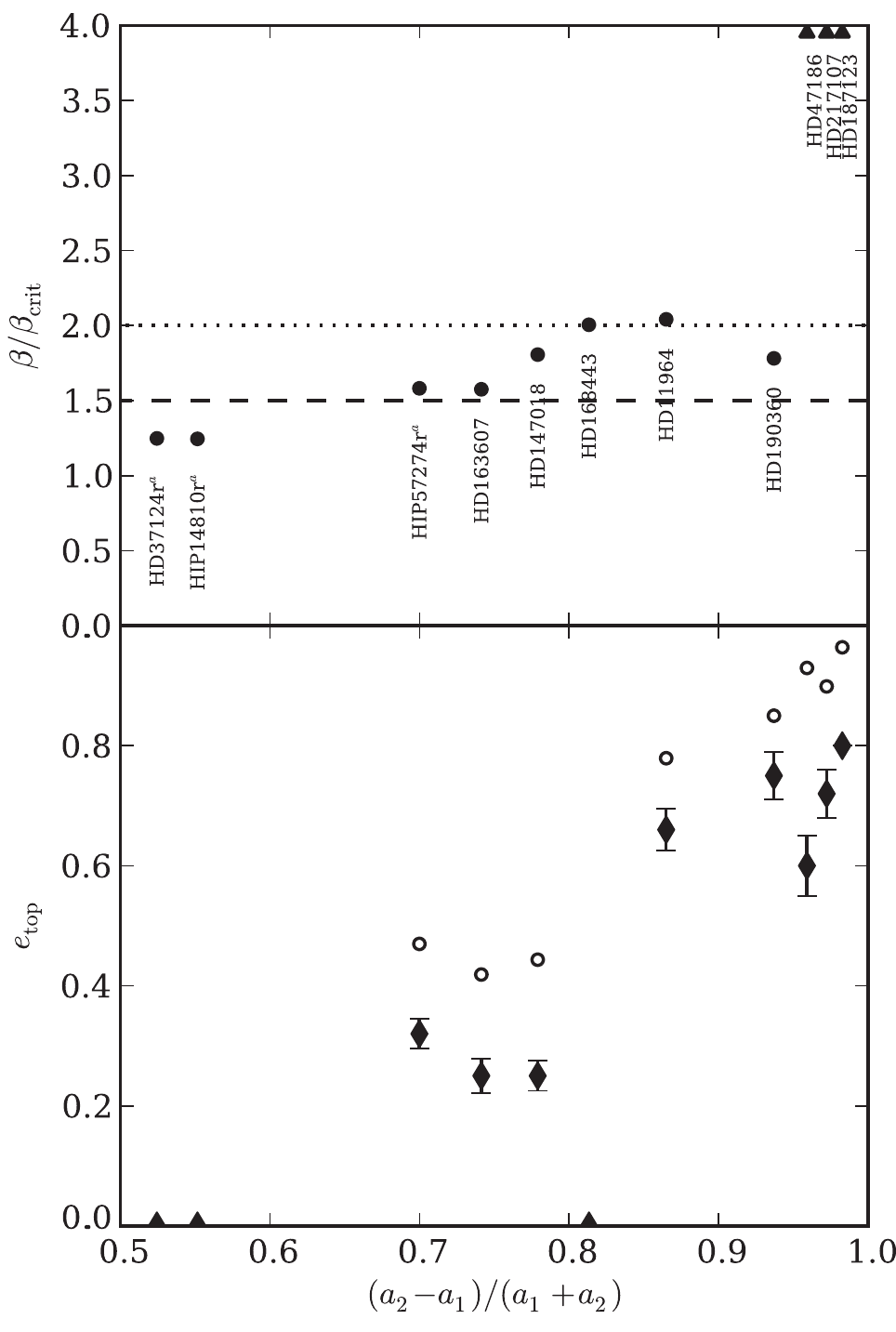}
\caption{Constraining stability zones analytically. Various measures of stability shown as a function of the normalized spacing of the enclosing planets. Systems with more than 2 known planets are marked with superscript $a$. \textit{Top panel}: Analytic stability criterion \bbcrit. The dashed lines indicates the minimum value for a system to enclose additional planets. Depending on the planet configuration, this line can shift up to 2.0 (dotted line). System above the dotted line always allow stable orbits. \textit{Bottom panel}: The maximum eccentricity $e_{\rm top}$. It is shown as circles when $c_i$ are obtained by fitting a piecewise curve to the data. The directly measured $e_{\rm top}$ from the simulations and their uncertainties are given as diamonds. } 
\label{fig:combi}%
\end{figure}

The top panel of figure \ref{fig:combi} shows \bbcrit of the planetary systems. In the case of the two systems that have the smallest separation in semi-major axis, they are well below \bbcrit$<1.5$. For the most separated systems, \bbcrit$>2.0$. In between, there are systems with $1.5<$\bbcrit$<2.0$ where the existence of additional enclosed stable orbits is not sure. The simulations show that in our sample, all system with \bbcrit$>1.5$ can harbor additional Super-Earths. Nevertheless, HD 168443 is right at the edge of \bbcrit$=2.0$ and only very few planets are stable.

The bottom panel of figure \ref{fig:combi} shows the maximum eccentricity $e_{\rm top}$ as a function of the separation. Systems containing planets with zero eccentricity would follow a straight line \citep[e.g.][]{Fang12} whereas high eccentricity planets with high masses are truncating the stable region, respectively $e_{\rm top}$, or even allow no stable region ($e_{\rm top}\leq0$). Large orbital spacing of the planets suppress this effect. We estimate $c_1$ and $c_2$ for every system separately. Then, calculating $e_{\rm top}$ results in an slight overestimation with respect to the maximum eccentricity observed directly in the simulations, because the piecewise function does not account correctly for the flatted top of the stable region. Hence, even if we would guess $c_1$ and $c_2$ correctly, we would overestimate slightly the hight of the stability zone with this analytic approach.

\subsection{Predicting habitable Super-Earths}

HD 168443 provides only very few stable simulations. Hence, we treat it as a fully packed system. All the systems we study in detail with massive test planets provide well defined regions with stable orbits for a $10 M_{\oplus}$ Super-Earth, partially located in the EHZ. We combine the stability of the orbits with the time-averaged location given by the occurrence rate, the analytic estimation of $e_{\rm forced}$ and the weak constraints from the FTD values. The location in the ($a,e$)-plane where a hypothetical Super-Earth is most likely to be observed is given in table \ref{tab:predictions} for each system.

\begin{table}
\centering
\caption{The most likely location in the (a,e)-plane for the observation of a hypothetical habitable Super-Earth. We comment on the system if there are features that could limit the habitability (high $e$) or the stability (small FTD, MMRs).} 
\begin{tabular}{l c c }
\hline
\input{predictions.txt}
\end{tabular}
\label{tab:predictions}
\end{table}

There exist predictions from previous studies. HD 47186 was studied in detail concerning the possible existence of a planet in the EHZ by \cite{Kopparapu08}. They found that a 10$\,M_\oplus$ planet is stable in the EHZ or even two 10$\,M_\oplus$ with low eccentricities can exist between planets $b$ and $c$. As mentioned above, we give a different estimate of the FTD map. The differences result from larger time steps used in the \cite{Kopparapu08} simulations. 

\cite{Gregory07} proposed that the planetary system HD 11964 consists of three instead of two planets based on fitting the Doppler spectroscopy data. Their three-planet solution is shown as green circles in figure \ref{fig:panel1} and is consistent with our stability region. The small difference in the orbital elements of the known planets would not significantly change the region. Nevertheless, the high eccentricity of the additional planet seems very unlikely and is outside of the EHZ. This three-planet solution was not confirmed by \cite{Wright09}. 

In HD 190360, \cite{Veras10} reported a stable terrestrial planet in the HZ might be possible according to test particle stability simulations, in agreement with our results.

In \cite{Jones06}, numerous systems are studied and the stability of a habitable Earth is estimated based on critical distances (basically parametrized by $c_1$ and $c_2$, see section \ref{sec:analytic}) to the giant planets. Hence, their estimation of the stability zone differs fundamentally from our fully numerical approach. Since in some system new planets were found in the meantime, we can only compare our results concerning HD 190360, HD 168443, HD 217107 and HD 37124. We agree on the survival of hypothetical planets in the HZ of HD 190360. We also found that stable orbits are unlikely in the HZ of HD 168443. In HD 217107, \cite{Jones06} localized the HZ at $2.0\lesssim a\lesssim 4.0$, which differs from our estimate. This results from the fact that we use slightly different stellar parameters and a newer estimation of the HZ \citep{Kopparapu13}. According to our results, the HZ is closer to the star and thus, the stability zone is partially located inside the HZ. In addition, \cite{Jones06} predicted that stable orbits can exist partially in the HZ of HD 37124. Out test particle simulations show that no additional planets are stable in the HZ and the analytic approach fails in this system. 

\subsection{Limited parameter space}

There are many parameters that control the architecture of a 2+1 planet system. Our simulations focus only on two dimensions (semi-major axis and eccentricity) of a multidimensional parameter space. Orbital inclinations, orbital phases and the mass of the test planet offer a wide range of additional scenarios to study. We think that only extreme values in inclination and mass will have a significant effect on the results: the orbital angles of the planets were chosen randomly in the simulations and only at the edges of the stable zone do these angles play any role regarding stability or FTD values. This could explain why the FTD does not always have a continuous gradient; meaning that sometimes small FTD values alternate with FTD $\approx1$ at the transition from high FTD to low FTD regions (for example HD 47186, $a\approx0.5$, $e>0.2$). Since massless and massive test planet simulations give very similar results, only test planets with masses $m\gg 10 M_{\oplus}$, small Neptune's, might put the stability of the system at risk.
Beside the parameters that control the orbit of the hypothetical planet, the orbital solution of the known planets is not unique. High inclination and masses can have a dramatic effect on the stability zone or on the stability of the known planets \citep{Veras10}.

Our simulations show that there are broad stable regions in many of the systems. These regions can harbor more than one Super-Earth size planet. But the parameter space increases rapidly by adding new planets, and we did not take this into account in additional simulations.

To test if the significance of our results depends on the simulation period of 10\,Myr, we carried out the simulations of HD 190360 for 50\,Myr. Indeed, we observed that the fraction of stable orbits reduces from 48.2 to 46.3 per cent. The overall shape and extension of the stability zone is not affected. Mostly, the additional unstable orbits are located at the MMRs which tent to stabilize orbits and the MMR are a bit more pronounced. All told, the limitation to 10\,Myr seems reasonable and does not influence our final results.

\section{Summary and conclusions}
\label{sec:conclusion}
We carry out numerous N-body simulations with the new GPU code GENGA to study the existence of hypothetical planets in extra-solar planetary systems. In nine systems, we study the stability of a 10$M_\oplus$ Super-Earth in high resolution in the ($a,e$)-plane. The reaction of the known planet on this hypothetical body and its movement in the ($a,e$)-parameter space allow us to predict the most likely orbital parameters of a Super-Earth in the habitable zone. Following the PPS-hypothesis, we find that for eight systems additional low mass planets can exist (apart from HD 168443), most of them with possible orbits in the EHZ (apart from HD 217107). The most promising candidate hosting a stable Super-Earth in its HZ with low $e$ is HD 11964 and, with a modest eccentricity of $e\approx 0.2$: HD 47186, HD 187123 and HD 190360.

Beside the lines of crossing orbits, MMRs with the outer planet are a main feature that shaped the stable region. Comparing the simulations with massless test particles and the simulations with massive planets, the main differences are found in the effect of the MMRs. While the $3:1$ MMR in HD 11964 results in a unstable wedge in the stable region, the same MMR is stable if the hypothetical planet is massive. 

Simulations in several systems show that close encounters are not a good criteria to identify unstable orbits. In some systems, a significant fraction of the planets on stable orbits are involved in such an energy exchange.

Beside the drawbacks of the FTD values, it does not constrain any of our stable zones in the HZ significantly (apart from HD 217107).  We point out that our FTD results concerning HD 47186 are fundamentally different to a previous study and shows some interesting secular resonance effects.

\section*{Acknowledgments}
We acknowledge the support of the Swiss National Science Foundation Grant No. 20020-127896 and thank the University of Zurich and HP$^2$C project for the financial support. The code was developed at CSCS and with HP$^2$C resources. The simulations were performed on zBox 4 supercomputer on NVidia GTX 590 graphic processor units at the University of Zurich. Thanks for the technical support to Doug Potter. We thank a anonymous reviewer for his helpful suggestions as well as Ravi Kopparapu and Ben Moore for useful discussions.
This research has made use of the Exoplanet Orbit Database and the Exoplanet Data Explorer at exoplanets.org.

\appendix
\section{Habitable Zone}
\cite*{Kasting93} (recently updated by \cite{Kopparapu13}) provide the inner and outer boundaries of the habitable zone (HZ). 
A planet on an eccentric orbit may partially escape from the habitable zone, even if its semi-major axis lies inside the HZ. \cite{Williams02} showed that orbit-average flux is the most important parameter for a long-term climate stability. 
The boundaries of the habitable zone around a star depends on its luminosity $L$ and its effective temperature $T_{\rm eff}$ as well as on planetary characteristics that control the greenhouse effect. The flux depends mainly on the luminosity. The effective temperature is a measure of the infrared fraction in $L$. A greater infrared fraction results in a greater greenhouse effect for a given stellar flux. Following the new estimates \cite{Kopparapu13}, the critical flux at the inner boundary of the HZ, where runaway greenhouse effect would take place and all surface water will evaporate and hydrogen will rapidly escape to space, is given by
\begin{equation}
\begin{split}
{S_i} = 1.0140+8.1774\times10^{-5} T_{\star}+1.7063\times10^{-9} T_\star^2 \\
-4.3241\times10^{-12} T_{\star}^3-6.6462\times10^{-16}T_{\star}^4,
\end{split}
\end{equation}
where $T_\star = T_{\rm eff}-5740 K$. The outer boundary flux corresponds to a minimum flux at which a maximum greenhouse effect can maintain liquid water on the surface of the planet with a cloud-free carbon dioxide atmosphere,
\begin{equation}
\begin{split}
{S_o} = 0.3483+5.8942\times10^{-5} T_{\star}+1.6558\times10^{-9} T_\star^2 \\
-3.0045\times10^{-12} T_{\star}^3-5.2983\times10^{-16}T_{\star}^4.
\end{split}
\end{equation}
The critical distances denoting the boundaries of the habitable zone are than given by the inverse square law:
\begin{equation}
\frac{r_i}{r_{\rm AU}} = \left(\frac{1}{S_i}\frac{L_\star}{L_\odot}\right)^{1/2},
\end{equation}
\begin{equation}
\frac{r_o}{r_{\rm AU}} = \left(\frac{1}{S_o}\frac{L_\star}{L_\odot}\right)^{1/2}.
\end{equation}
$L_\odot$ is the solar luminosity and $L_\star=4\pi R_\star\sigma T_{\rm eff}$ is the luminosity of the star, a function of the radius of the star, $R_\star$. $r_{\rm AU}$ denotes the distance of Sun and Earth.

We focus on planets which receive as much flux over one orbit as a planet on circular orbit with the same semi-major axis confined in the HZ, we have to take into account the eccentricity dependent orbit-averaged mean flux \citep{Williams02,Adams06}:
\begin{equation}
\langle F \rangle = \frac{F}{4\pi a^2\sqrt{1-e^2}}.
\end{equation}
Hence, we assume that his flux corresponds to the critical fluxes at the HZ boundaries for e=0 and we can deduce constraints for an orbit with elements ($a,e$) inside these boundaries:
\begin{equation}
r_i<a(1-e^2)^{1/4}<r_o.
\label{eq:HZparas}
\end{equation}
We will refer to this concept of the HZ as the eccentric HZ (EHZ) \citep{Barnes08,Kopparapu08}.

\bibliographystyle{mn2e_new}
\bibliography{mybib}
\end{document}

%% file: planetlist.txt
Star	&	$M_\star$ [$M_{\odot}$]	&	$T_\star$ [K]	&	$R_\star$ [$R_\odot$]	&	Planet	&	$m \sin i$ [$M_{\rm jup}$]	&	$a$ [AU]	&	$e$		\\
\hline																	
HIP 14810	&	0.99	&	5485	&	1.32	&	b	&	3.874	&	0.0692 $\pm$ 0.00115	&	0.14248 $\pm$ 0.00095			\\
	&		&		&		&	c	&	1.275	&	0.5454 $\pm$ 0.0091	&	0.153 $\pm$ 0.0132			\\
	&		&		&		&	d	&	0.581	&	1.886 $\pm$ 0.036	&	0.165 $\pm$ 0.04			\\
HD 37124 	&	0.85	&	5500	&	0.77	&	b	&	0.674	&	0.5336 $\pm$ 0.0089	&	0.054 $\pm$ 0.028			\\
	&		&		&		&	c	&	0.648	&	1.710 $\pm$ 0.029	&	0.125 $\pm$ 0.055			\\
	&		&		&		&	d	&	0.687	&	2.807 $\pm$ 0.06	&	0.16 $\pm$ 0.14			\\
HD 163607 	&	1.09	&	5543	&	1.7	&	b	&	0.769	&	0.3592 $\pm$ 0.006	&	0.730 $\pm$ 0.02		\\
	&		&		&		&	c	&	2.292	&	2.418 $\pm$ 0.041	&	0.120 $\pm$ 0.06			\\
HIP 57274 	&	0.73	&	4640	&	0.68	&	b	&	0.037	&	0.0713 $\pm$ 0.00163	&	0.19 $\pm$ 0.1			\\
	&		&		&		&	c	&	0.41	&	0.1778 $\pm$ 0.0041	&	0.050 $\pm$ 0.02			\\
	&		&		&		&	d	&	0.529	&	1.007 $\pm$ 0.027	&	0.270 $\pm$ 0.05			\\
HD 190360 	&	0.983	&	5552	&	1.08	&	b	&	1.535	&	3.973 $\pm$ 0.071	&	0.313 $\pm$ 0.0191		\\
	&		&		&		&	c	&	0.059	&	0.1292 $\pm$ 0.0022	&	0.237 $\pm$ 0.082			\\
HD 147018	&	0.927	&	5441	&	1.053	&	b	&	2.127	&	0.2389 $\pm$ 0.004	&	0.4686 $\pm$ 0.0081			\\
	&		&		&		&	c	&	6.593	&	1.923 $\pm$ 0.039	&	0.133 $\pm$ 0.011			\\
HD 168443 	&	0.995	&	5491	&	1.59	&	b	&	7.697	&	0.2938 $\pm$ 0.0049	&	0.529 $\pm$ 0.024			\\
	&		&		&		&	c	&	17.386	&	2.853 $\pm$ 0.048	&	0.2113 $\pm$ 0.00171			\\
HD 11964 	&	1.107	&	5349	&	1.67	&	b	&	0.618	&	3.155 $\pm$ 0.059	&	0.041 +0.088/-0		\\
	&		&		&		&	c	&	0.078	&	0.2285 $\pm$ 0.0038	&	0.30 $\pm$ 0.17			\\
HD 47186 	&	0.99	&	5675	&	1.13	&	b	&	0.071	&	0.04984 $\pm$ 0.00083	&	0.038 $\pm$ 0.02		\\
	&		&		&		&	c	&	0.348	&	2.387 $\pm$ 0.078	&	0.249 $\pm$ 0.073			\\
HD 217107 	&	1.108	&	5704	&	1.5	&	b	&	1.401	&	0.0750 $\pm$ 0.00125	&	0.1267 $\pm$ 0.0052		\\
	&		&		&		&	c	&	2.615	&	5.33 $\pm$ 0.2	&	0.517 $\pm$ 0.033			\\
HD 187123 	&	1.037	&	5815	&	1.14	&	b	&	0.51	&	0.04209 $\pm$ 0.0007	&	0.0103 $\pm$ 0.0059		\\
	&		&		&		&	c	&	1.942	&	4.83 $\pm$ 0.37	&	0.252 $\pm$ 0.033			\\

%% file: bvalues.txt
Star	& pair &	\bbcrit	\\
\hline																	
HIP 14810$^{\rm a}$	    & c-d &	1.245	\\
HD 37124$^{\rm a}$ 	    & b-c &	1.248	\\
HD 163607 	            & b-c &	1.575	\\
HIP 57274$^{\rm a}$ 	& c-d &	1.581	\\
HD 190360 	            & b-c &	1.781	\\
HD 147018	         	& b-c &	1.806	\\
HD 168443 	            & b-c &	2.005	\\
HD 11964 	            & b-c &	2.041	\\
HD 47186 	        	& b-c &	6.134	\\
HD 217107 	            & b-c &	8.941	\\
HD 187123 	            & b-c &	15.091	\\

%% file: percentages.txt
system	& $\Delta t$ [d] & $\mathcal O$ & $a_{\rm min}$ & $a_{\rm max}$ & $e_{\rm max}$ & $N_{\rm init}$ & $N_{\rm stab}$ & $N_{\rm Hill}$  & $f_{\rm stab} (\%)$ & $F_{\rm stab} (\%)$ & $F_{\rm stab,m=0} (\%)$ 	\\
\hline
HD 163607   & 0.80    & 4 &  0.50    & 2.0 & 0.5 & 5000 & 680    & 24  & 13.6      & 4.9     & 5.5   \\
HD 217107   & 0.35    & 4 &  0.10    & 3.0 & 0.8 & 5000 & 1883   & 0   & 39.5      & 17.3    & 16.4  \\
HIP 57274   & 0.40    & 2 &  0.20    & 0.7 & 0.5 & 5000 & 1149   & 1   & 23.2      & 6.2     & 5.7   \\
HD 11964    & 2.00    & 4 &  0.25    & 2.8 & 0.7 & 5000 & 2986   & 129 & 61.4      & 37.4    & 34.8  \\
HD 187123   & 0.50    & 2 &  0.20    & 3.5 & 0.8 & 5000 & 2891   & 30  & 56.0      & 34.0    & 32.6  \\
HD 147018   & 1.00    & 4 &  0.30    & 1.3 & 0.5 & 5000 & 729    & 113 & 15.0      & 4.4     & 5.3   \\
HD 47186    & 0.50    & 4 &  0.06    & 2.3 & 0.65 & 5000 & 2205   & 122 & 55.9      & 35.1    & 32.4  \\
HD 168443   & 0.30    & 2 &  0.70    & 1.5 & 0.4 & 5000 & 49     & 48  & 0.9       & 0.1     & 1.1    \\
HD 190360   & 0.85    & 2 &  0.15    & 2.5 & 0.8 & 5000 & 2409   & 11  & 48.2      & 23.5    & 23.0   \\

%% file: predictions.txt
system	&  	stable region in HZ ($a,e$) & comment\\
\hline
HD 11964    &  (1.3-2.4 AU, 0.05)       & - \\
HD 47186    &  (0.9-1.3 AU, 0.1-0.3)    & high $e$ \\
HD 147018   &  (0.8-0.9 AU, 0.0-0.1)    & - \\
HD 163607   &  (1.3-1.4 AU, 0.05-0.1)   & - \\ 
HD 187123   &  (1.0-2.2 AU, 0.1-0.3)    & high $e$ \\
HD 190360   &  (0.8-1.5 AU, 0.1-0.3)    & high $e$ \\
HD 217107   &  (1.3-1.6 AU, 0.3)        & small FTD\\
HIP 57274   &  (0.37-0.56 AU, 0.1-0.3)  & strong MMRs\\